\documentclass[prd,preprint,superscriptaddress,nofootinbib]{revtex4}
\usepackage{graphicx}
\usepackage{amsfonts}
\usepackage{latexsym}
\usepackage{amssymb}
\usepackage{amsmath}
\usepackage{epsfig}
\usepackage{color}

\begin{document}

\title{Relaxion window}

\author{Tatsuo Kobayashi}
 \affiliation{Department of Physics, Hokkaido University, Sapporo 060-0810, Japan
}

\author{Osamu Seto}
 \affiliation{Department of Life Science and Technology,
  Hokkai-Gakuen University, Sapporo 062-8605, Japan}
 \affiliation{Office of International Affairs, Hokkaido University, Sapporo 060-0815, Japan}
 \affiliation{Department of Physics, Graduate School of Science, Hokkaido University, Sapporo 060-0810, Japan}

\author{Takashi Shimomura}
 \affiliation{Faculty of Education, University of Miyazaki,
 Miyazaki, 889-2192, Japan
}

\author{Yuko Urakawa}
 \affiliation{Department of Physics and Astrophysics, Nagoya University, Nagoya 464-8602, Japan
}

%
\begin{abstract}
We investigate cosmological constraints on the original relaxion scenario proposed
 by Graham, Kaplan and Rajendran. 
We first discuss the appropriate sign choice of the terms in the
 scalar potential, when the QCD axion is the relaxion with a relaxion-inflaton coupling proposed in the original paper.
We next derive the cosmologically consistent ranges of the mass and a coupling of
 the relaxion for both the QCD relaxion and non-QCD relaxion. 
The mass range is obtained by $10^{-5}$ eV $\ll m_{\phi} \lesssim 10^4$ eV. 
We also find that
 a strong correlation between the Hubble parameter at the relaxion stabilization
 and the scale $\Lambda$ of non-QCD strong dynamics, which generates
 the non-perturbative relaxion cosine potential. For a higher
 relaxion mass, a large scale $\Lambda$ becomes available. However, for
 its lower mass, $\Lambda$ should be small and constructing such a
 particle physics model is challenging. 
\end{abstract}

\preprint{EPHOU-16-007}
\preprint{HGU-CAP-041} 
\preprint{UME-PP-004}

\vspace*{3cm}
\maketitle


\section{Introduction}

Graham, Kaplan and Rajendran recently proposed a radical solution to the electroweak
 hierarchy problem by the dynamics of a scalar field~\cite{Graham:2015cka}. 
The basic idea had been
 proposed in the context of the cosmological constant problem~\cite{Abbott:1984qf}.
During inflation~\cite{Inflation1,Inflation2,Inflation3,Inflation4}, the axion-like scalar field called the relaxion $\phi$
 slowly rolls down the potential and reduces, with the coupling to the standard model (SM) Higgs field $\Phi$,
 the effective Higgs mass term from a huge positive value with the relaxion-Higgs coupling, $ - g \phi|\Phi|^2 $.
The Higgs field is expected to have the cut-off scale size mass term $M^2|\Phi|^2$
 with $M$ being the cutoff scale of the theory.
When the relaxion field goes across the field value $M^2/g$ 
 and the effective Higgs mass slightly becomes tachyonic,
 the electroweak symmetry breaking takes place.
The relaxion field is fixed by a non-perturbative effect induced potential
 at $\phi\simeq M^2/g$, that
 stabilizes the vacuum expectation value (VEV) and
 the mass of the electroweak scale for the SM Higgs field.
As a result of the substantial cancellation between $M^2$ and
 $g\phi$ in the SM Higgs mass term, the lower electroweak scale
 compared to the cut off scale $M$ can be realized. 

Since the proposal~\cite{Graham:2015cka},
 some aspects of the relaxion scenario and various extended models have been
 studied~\cite{Espinosa:2015eda,Hardy:2015laa,Patil:2015oxa,
 Jaeckel:2015txa,Gupta:2015uea,Batell:2015fma,Matsedonskyi:2015xta,Marzola:2015dia,Choi:2015fiu,DiChiara:2015euo,Ibanez:2015fcv,Fonseca:2016eoo,Fowlie:2016jlx,Evans:2016htp,Huang:2016dhp}.
For instance, the origin of non-perturbative terms~\cite{Espinosa:2015eda,Matsedonskyi:2015xta},
 fine-tuning~\cite{Jaeckel:2015txa,Gupta:2015uea,DiChiara:2015euo,Fowlie:2016jlx},
 supersymmetric extension~\cite{Batell:2015fma,Evans:2016htp},
 the issue of large excursion of relaxion field~\cite{Choi:2015fiu,Ibanez:2015fcv,Fonseca:2016eoo}
 have been considered.

In this paper, in the context of the original relaxion model proposed in Ref.~\cite{Graham:2015cka},
 we examine a consistent parameter region from a cosmological viewpoint. 
Namely, we examine the dynamics of original models during and after inflation. 
This paper is organized as follows. 
In section~\ref{review}, we briefly review the outline of the model proposed in Ref.~\cite{Graham:2015cka}. 
In section~\ref{QCD}, we study inflationary dynamics of the model, introducing the relaxion-inflaton coupling.
In Ref.~\cite{Graham:2015cka}, it was suggested that with this coupling,
 the QCD axion can serve the relaxion, evading the strong CP problem in the SM
 if parameters are appropriately taken.
We examine this possibility with a special attention to the
 sign of each term in the scalar potential, which may not be very
 clear in the original paper~\cite{Graham:2015cka}.
In section~\ref{nonQCD}, 
 we study dynamics of relaxion and Higgs transition during inflation and
 calculate the relic abundance of relaxion generated as coherent
 oscillation.  
We discuss the conditions that the relaxion transition
 does not disturb the inflationary cosmic expansion and show that these
 conditions constrain the parameter space of the relaxion.
We also find that this abundance does not exceed the dark matter (DM) abundance if 
 a relaxion potential height is low enough or inflation continues during
 certain number of $e$-fold after the relaxion stabilization.
From the experimental search limit
 for axion-like particles (ALPs) and the consistency of the scenario, 
 we discuss the viable parameter space of the relaxion model and summarize it
 in plots of the relaxion mass and coupling plane.
Section \ref{conclusion} is devoted to concluding remarks.

\section{Brief review of relaxion scenario}  
\label{review}

Here, we briefly describe the outline of relaxion mechanism based on
 the original minimal scenario proposed in Ref.~\cite{Graham:2015cka}.
The model contains three scalar fields; the SM Higgs $\Phi$, the
 relaxion $\phi$ and the inflaton $\sigma$.
The scalar potential for $\Phi$ and $\phi$ is given by 
\begin{equation}
V(\Phi,\phi) = \pm  (M^2 - g \phi)|\Phi|^2+\lambda|\Phi|^4 \pm (-g M^2 \phi +\cdots) ,
\label{potential}
\end{equation}
 where $g$ is a soft-breaking parameter of the periodic shift symmetry
 of the relaxion $\phi$. Here, we also wrote the
 Higgs self-interaction term with the coupling $\lambda$ which 
 was omitted in Ref.~\cite{Graham:2015cka}.  
Since the effective mass term should be
 positive  before the symmetry breaking, the initial value of $\phi$
 should be chosen such that $\pm (M^2 - g \phi) > 0$. 
At $\phi = \phi_c  \equiv M^2/g$, the mass term vanishes and afterwards
the Higgs field becomes tachyonic. Then, 
the Higgs field develops a non-vanishing VEV. To make this possible, we
need to choose the sign of the last term as in Eq.~(\ref{potential}).
For example, when we choose the mass term as $-(M^2 - g \phi)|\Phi|^2$, the initial
 value of $\phi$ should be largely positive with $\phi > M^2/g$. This
 requires the sign of the last term to be positive, i.e., $+g M^2 \phi$. 
We will find that the symmetry breaking does not take place during the
slow-roll inflation in the other choice with $+(M^2 - g \phi)|\Phi|^2$.

Once $\Phi$ starts to take a non-vanishing field value,
 the relaxion field acquires a periodic potential
\begin{equation}
  \delta V(\phi) = m_\phi^2 f^2 \left( 1 - \cos \left( \frac{\phi}{f} \right) \right) ,
\label{pot:cos}
\end{equation}
 with $f$ being the relaxion decay constant,
 induced by non-perturbative effects in a strong interaction 
\begin{equation}
\delta {\cal L} = \frac{g_s^2}{32\pi^2}\frac{\phi}{f}G\tilde{G} ,
\end{equation}
 where $g_s$ is its gauge coupling and,
 $G$ and $\tilde{G}$ denote the field strength and its dual
 respectively. 
If we identify this strong interaction with the SM QCD, the height of the potential
\begin{equation}
 m_{\phi}^2 f^2 \sim (0.01 \, {\rm GeV}^2)^2 ,
\end{equation}
 is proportional to the VEV of the Higgs field, i.e.,
\begin{equation}
  m_{\phi}^2 f^2 \propto \Phi .
 \label{eq:Lambda-Phi}
\end{equation}
In order to show the $\Phi$ dependence of the mass explicitely,
 we introduce $\overline{m}_{\phi}$ defined by $ m_{\phi}^2 =\overline{m}_{\phi}^2 \frac{|\Phi|}{v/\sqrt{2}} $
 and rewrite the scalar potential (\ref{pot:cos}) as
\begin{equation}
  \delta V(\Phi,\phi) = \overline{m}_{\phi}^2 f^2 \frac{|\Phi|}{v/\sqrt{2}} \left( 1 - \cos \left( \frac{\phi}{f} \right) \right) ,
\label{pot:cos2}
\end{equation}
 with $v\simeq246$ GeV.

With a certain choice of the initial condition of $\phi$, the effective
Higgs mass squared $(M^2-g\phi)$ stays positive at an early stage of
inflation. During inflation, $\phi$ slowly rolls down the potential. 
When $\phi$ passes the field value $M^2/g$, the effective Higgs mass
becomes tachyonic and finally it can be settled down at a scale
 which is much smaller than the cutoff scale $M$.
This is the mechanism to realize the small weak scale, starting from the cutoff scale.
As the VEV of the Higgs field increases,
 the height of the cosine potential increases and $\delta V$ becomes
 more important (see Eq.~(\ref{eq:Lambda-Phi})).

The relaxion stops rolling down the potential, when the gradient of
the potential $-gM^2 \phi$ becomes comparable to the one of $\delta V$,
i.e., $gM^2 \simeq m^2_\phi f$. Solving the stationary condition for the total potential $
V(\Phi,\phi) +\delta V(\phi)$, we can estimate (the order of) the final
value of the relaxion as~\footnote{Here, for simplicity, we neglect the $\phi$ dependence of
$m^2_{\phi} f^2$ through the VEV of the Higgs field.}
\begin{equation} 
 \langle\phi\rangle \simeq f \arcsin\left( \frac{g M^2}{ m_{\phi}^2 f}\right).
\label{phi:min}
\end{equation}
This implies that the $\theta$ parameter,
 given by $\theta = \langle\phi\rangle/f$, is of order unity
 and leaves the strong CP problem unsolved. 
Therefore, in this minimum setup, we can not identify $\phi$ with the QCD axion. 
In order to circumvent the difficulty, in Ref.~\cite{Graham:2015cka},
 two possible scenarios were considered: an introduction of a
 relaxion-inflaton coupling and a non-QCD axion. 
In the following two sections, we discuss dynamics and constraints for both the QCD relaxion
 and a non-QCD based model.

\section{Dynamics of the QCD relaxion}  
\label{QCD}

In this section, we discuss the time evolution of the coupled
system with the SM Higgs $\Phi$, the
relaxion field $\phi$, and the inflaton $\sigma$ during inflation. We
postulate the ansatz of the total scalar potential as
\begin{equation}
 V(\Phi,\phi,\sigma) = V(\sigma) + V(\Phi,\phi) + \delta V(\Phi,\phi) + \kappa \sigma^2 \phi^2,
\label{fullscalarpotential}
\end{equation}
where additionally we introduced the coupling between the inflaton and the relaxion. 
The explicit forms of $V(\Phi,\, \phi)$ and $\delta V(\Phi,\phi)$ are given in the previous section. 
Notice that the sgins of $V(\Phi,\, \phi)$ given in Eq.~(\ref{potential})
 is same as Eq.~(1) in the original paper~\cite{Graham:2015cka}. 
As one will see, those signs lead to crucial difference. 
In our discussion, we do not specify the inflaton potential $V(\sigma)$.

One may expect that if the effective value of the gradient of the $\phi$ potential,
 $g M^2$, which should be comparable to $m^2_\phi f$ until around the relaxion stabilization,
 is significantly reduced after the stabilization, the relaxion still can be identified with the QCD
axion, evading the strong CP problem. 
To study this possibility, the last term in the scalar potential (\ref{fullscalarpotential})
 was introduced in Ref.~\cite{Graham:2015cka}. 
For the inflaton potential to be bounded from below, we assume that $\kappa$ takes a positive
 value.~\footnote{Introducing an inflaton potential $\sigma^p$ where $p$
is an even number with $p \geq 4$,
 we can keep the inflaton potential bounded from below also for $\kappa<0$. 
To fulfill the condition $ g^2 \ll \kappa \sigma^2$,
 a large field evolution may be preferred. 
However, a large field evolution where the $\sigma^p$ term dominates the inflaton potential
 is strongly disfavored by data of the cosmic microwave background anisotropies~\cite{Ade:2015lrj}.}

First, we consider the case where the mass term is given by $+(M^2- g \phi) |\Phi|^2$. 
Then, the effective slope of the relaxion during inflation is given by 
 $-g M^2 + 2 \kappa \sigma^2 \phi $. 
We assume that during inflation this slope is
 predominantly sustained by the coupling between the inflaton and the
 relaxion, i.e., $g M^2 \ll \kappa \sigma^2 \phi \simeq \kappa\sigma^2 M^2/g$,
which gives the lower bound on the value of $\sigma$ as
$$ g^2 \ll \kappa \sigma^2.$$ One may naively expect that the relaxion can
 be stabilized,
 when the non-perturbative term $\delta V$ becomes comparable
 to the effectively enhanced slope $2 \kappa \sigma^2 \phi$.
Then, since the relaxion stabilization does not impose any constraint on $g
 M^2$, we can choose the value so that the small QCD $\theta$ term can
 be realized.

However, a little more careful consideration reveals that it is rather
difficult to stabilize the relaxion in this setup. The relaxion $\phi$
is supposed to evolve from a smaller value to a larger value, reducing
the effective Higgs mass squared from the huge positive value to a negative value. 
In the presence of the coupling with the inflaton, the relaxion
potential, given by
\begin{equation}
 V = -g M^2 \phi + \kappa \sigma^2 \phi^2  + \cdots , 
\end{equation}
has the minimum at $\phi_\textrm{min}=gM^2/(2\kappa \sigma^2)$. Using 
$ g^2 \ll \kappa\sigma^2$, we find $\phi_c \gg \phi_\textrm{min}$. 
Therefore, for the relaxion to be settled down at
$\phi=\phi_c$, $\phi$ should overshoot the potential minimum and climb
up the potential.  This is impossible, since we start with a small kinetic
 energy of $\phi$, which leads to the slow-roll evolution.  

In order to avoid this problem, the sign of the terms in the scalar potential should be
\begin{equation}
V(\Phi,\phi) = -(M^2 - g \phi)|\Phi|^2+\lambda|\Phi|^4 - (-g M^2 \phi +\cdots),
\label{QCDchoice}
\end{equation}
where the bare Higgs mass squared is large negative and $\phi$ evolves from a large positive to negative.
Thus, the introduction of the $\kappa\sigma^2\phi^2$ lets
 the QCD axion play the role of the relaxion,
 only if the sign of the terms are appropriately chosen.

\section{Parameter region}     \label{nonQCD}

In this section, we discuss the allowed parameter range for the non-QCD relaxion model. 
The dynamical scale is given by $\Lambda^2 = m_\phi f$.
Also extra matter fields, which couple with the Higgs field $\Phi$, are 
 required in order to have a $\Phi$-dependence in such a potential like (\ref{eq:Lambda-Phi}).
Notice that the same argument also can apply to the QCD relaxion model,
 where the gradient of the relaxion potential has inflaton dependence and
 changes from the one at the stabilization time after inflation,
 if the signature of the divergent mass term could be controlled as in Eq.~(\ref{QCDchoice}).

\subsection{Relaxion dynamics and stability condition}
\label{dynamics}

In this subsection, we derive a condition that the relaxion is
stabilized at $\phi \simeq \phi_c = M^2/g$. Until the time $t=t_{\Phi}$
when the relaxion passes the field value $\phi = M^2/g$ and the Higgs
field acquires the tachyonic mass at the origin, $\phi$ and $\sigma$
slowly rolls down the potential. Under the slow-roll approximation, the field equations
are given by 
\begin{eqnarray}
 && H_-^2 \simeq \frac{1}{3M_P^2}V(\Phi,\phi,\sigma) , \\
 && 3H_- \dot{\sigma} + \partial_{\sigma}V(\sigma) \simeq 0, \\
 && 3H_- \dot{\phi} + \partial_{\phi}(V(\Phi,\phi) + \delta V(\Phi,\phi) ) \simeq 0 ,
\end{eqnarray}
where $M_P$ denotes the reduced Planck mass, $M_P \simeq 2.4 \times 10^{18}$ GeV. 
We assume that the transition at $t_{\Phi}$ does not disturb the
inflationary dynamics, imposing 
\begin{eqnarray}
 3 M_P^2 H_-^2 \simeq  3 M_P^2 H_+^2 \gg {\rm Max} \left[ {\cal O}(v^4), {\cal O}(m_{\phi}^2 f^2) \right] .
 \label{cond:smoothinflation}
\end{eqnarray}
Here $H_-$ denotes the Hubble parameter for $t < t_{\Phi}$ 
and $H_+$ denotes the one for $t> t_{\Phi}$.
If one of those conditions is violated,
 the scenario suffers from either too strong temperature effect
 or too abundant relaxion energy density, as we will show in the next subsection.

After $t_{\Phi}$, the relaxion is not necessarily in the slow
 roll phase anymore
 and the field equations for $\Phi$ and $\phi$ are given by
\begin{eqnarray}
 &&  \ddot{\Phi}+ 3H_+ \dot{\Phi} + \partial_{\Phi}(V(\Phi,\phi) + \delta V(\Phi,\phi) ) =0, \\
 &&  \ddot{\phi}+ 3H_+ \dot{\phi} + \partial_{\phi}(V(\Phi,\phi) + \delta V(\Phi,\phi) ) =0 .
\end{eqnarray}
The potential for the relaxion is now given by
\begin{align}
 V(\Phi,\phi) + \delta V(\Phi,\phi) = & (M^2 - g \phi)|\Phi|^2+\lambda|\Phi|^4 -g M^2 \phi \nonumber \\
 & + \overline{m}_\phi^2 f^2 \, \frac{|\Phi|}{v/\sqrt{2}} \left( 1 - \cos \left( \frac{\phi}{f} \right) \right) + \cdots\, .
\end{align}

The values of $\Phi$ and $\phi$ at a local minimum, $\Phi = (0,
v_h/\sqrt{2})$ and $\phi=v_{\phi}$, can be determined by solving
\begin{eqnarray}
 &&  v_h\left( M^2-g v_{\phi} + \lambda v_h^2 \right)
 +\frac{\overline{m}_{\phi}^2 f^2}{v}\left( 1-\cos \left(\frac{v_{\phi}}{f}\right) \right) =0,\\
 && -g\left( M^2+\frac{v_h^2}{2}  \right) 
 +\frac{\overline{m}_{\phi}^2 f v_h}{v}\sin\left(\frac{v_{\phi}}{f}\right)  = 0 .
\end{eqnarray}
When the strong dynamical scale is smaller than the weak scale,
i.e., $v^2 \gtrsim \overline{m}_{\phi} f$, we can further rewrite these conditions as
\begin{eqnarray}
 &&  v_h\left( M^2-g v_{\phi} + \lambda v_h^2 \right) \simeq 0,
\label{stat:h} \\
 && -g M^2 
 +\frac{\overline{m}_{\phi}^2 f v_h}{v}\sin\left(\frac{v_{\phi}}{f}\right) \simeq 0,
 \label{stat:phi}
\end{eqnarray}
 and obtain the resultant minimum field values as
\begin{eqnarray}
 v_h \sim  \sqrt{\frac{-(M^2-g v_{\phi})} {\lambda}} \sim  v, 
\label{min:vh_inf} 
\end{eqnarray}
and 
\begin{eqnarray}
 v_{\phi} & \sim & \frac{M^2}{g} \label{min:vphi_inf1}  \\
         & \sim & f \arcsin\left( \frac{g M^2}{ \overline{m}_{\phi}^2 f}\right) \qquad (\textrm{mod} \,\,\, 2\pi) ,
\label{min:vphi_inf2} 
\end{eqnarray}
 where (\ref{min:vphi_inf1}) and (\ref{min:vphi_inf2}) are obtained
 from Eqs.~(\ref{stat:h}) with $|M^2-g v_{\phi}| \ll M^2 $ and (\ref{stat:phi}), respectively.
The final value of $v_h$ is expected to be $v$ to reproduce the observed Fermi constant.
We find that, from Eqs.~(\ref{stat:phi}), (\ref{min:vh_inf}) and (\ref{min:vphi_inf2}),
 the condition for relaxion potential has a local minimum is
\begin{equation}
 \frac{ g M^2 }{\overline{m}_{\phi}^2 f} < 1 .
\label{cond:steep}
\end{equation}
It should be noted that Eqs.~(\ref{min:vphi_inf1}) and (\ref{min:vphi_inf2}) are consistent only when 
$g/\overline{m}_{\phi} < 1$ is satisfied.

The effective mass matrix for the Higgs boson and relaxion is given as
\begin{eqnarray}
\begin{pmatrix}
M^2-g v_{\phi} + 3\lambda v^2 & -gv+\frac{\overline{m}_{\phi}^2 f^2}{f v}\sin\left(\frac{v_{\phi}}{f}\right) \\
-gv+\frac{\overline{m}_{\phi}^2 f^2}{f v}\sin\left(\frac{v_{\phi}}{f}\right)
 & \overline{m}_{\phi}^2\frac{v_h}{v}\cos\left(\frac{v_{\phi}}{f}\right) 
\end{pmatrix} .
\label{mass:h-a}
\end{eqnarray}
If 
\begin{equation}
\overline{m}_{\phi}^2\frac{v_h}{v}\cos\left(\frac{v_{\phi}}{f}\right) \simeq m_{\phi}^2 > H_+^2
\label{cond:axion_stabil2}
\end{equation}
holds, the relaxion-like direction can be stabilized during
inflation. 
This condition requires that the relaxion should be
stabilized kinematically and has not been discussed
elsewhere~\footnote{In this paper, we do not impose Eq.~(7) in Ref.~\cite{Graham:2015cka}. This is because it is subtle if ``the classical beats quantum
condition'' can be actually given by their (7) even in the single field
model (see, e.g., Ref.~\cite{Vennin:2015hra}). This condition will become more
complicated in the presence of multiields like in the relaxion
scenario.}. 
When Eq.~(\ref{stat:phi}) is fulfilled, the mass matrix can be rewritten as 
\begin{eqnarray}
\begin{pmatrix}
 m_h^2 & \frac{gM^2}{v} \\
\frac{gM^2}{v} & m_{\phi}^2
\end{pmatrix}\,,
\end{eqnarray}
and the Higgs-relaxion mixing is given by
\begin{equation}
 \tan 2 \theta_{h\phi} = \frac{2gM^2}{v(m_h^2-m_\phi^2)}.
\label{mix:h-phi}
\end{equation}

\subsection{Relaxion relic abundance}

The relaxion starts to oscillate around the local minimum $v_\phi$
with the initial amplitude $\Delta\phi \sim f$ at $t = t_{\rm osc}$
during inflation. When the potential of the relaxion expanded around
$v_\phi$ is (approximately) quadratic,
 the energy density of the relaxion at the end of inflation is given by
\begin{equation}
\rho_{\phi}(t_{\rm end}) = \frac{1}{2}m_{\phi}^2 (\Delta\phi)^2
  \left( \frac{a(t_{\rm osc})}{a(t_{\rm end})} \right)^{3}.
\end{equation}
After the reheating by inflaton decay, the ratio between the energy
density of the relaxion and the entropy is given by
\begin{equation}
\frac{\rho_{\phi}}{s} 
 \simeq \frac{T_R}{4} \frac{m_{\phi}^2 (\Delta\phi)^2}{2 M_P^2 H_+^2}
 e^{-3 N_{\rm osc}} ,
\end{equation}
where we assumed that the energy density of the oscillating inflaton decreases as $a^{-3}$.
  Here, we used $
N_{\rm osc} \equiv \ln (a(t_{\rm end})/a(t_{\rm osc}))$. 
This can be rewritten as 
\begin{equation}
 \Omega_{\phi} h^2 \simeq 0.9 \times 10^9 \frac{T_R}{4\, {\rm GeV}}
 \left(\frac{m_{\phi}^2 (\Delta\phi)^2}{6 M_P^2 H_+^2} \right) e^{-3 N_{\rm osc}}  .
\end{equation}
When the lifetime of $\phi$ is longer than the age of our Universe,
 $\Omega_{\phi} h^2$ should not exceed
 the present dark matter density $\Omega_{CDM} h^2 \simeq 0.12$~\cite{Ade:2015xua}.
The cosmological dark matter abundance constrain model parameters as 
\begin{equation}
\Omega_{\phi} h^2 \lesssim 0.12 . 
\label{cond:omegah2-phi}
\end{equation}
If this constraint is barely satisfied, then the relaxion plays a role of dark matter. 
This is realized if $N_{\rm osc}$ is non-negligible or a value inside the bracket is very small,
 namely $ m_{\phi}^2 f^2 \sim m_{\phi}^2 (\Delta\phi)^2 \ll M_P^2 H_+^2$. 

For this relaxion mechanism to work, we need to keep the relaxion and
the Higgs field stabilized at the minimum after inflation. In particular, for the Higgs field, the temperature of
the thermal plasma should not be higher than of the weak scale $\sim v$. 
During the period between the end of inflation and the completion of reheating, 
the radiation has been generated by a partial decay of the inflaton $\sigma$ and 
the temperature is given by~\cite{KolbTurner}
\begin{equation}
 T \simeq \left( \sqrt{\frac{90}{\pi^2 g_*(T_R)}} T_R^2 M_P H \right)^{1/4} ,
\label{eq:T-HrelationInMD}
\end{equation}
 where $T_R$ is the reheating temperature, which has to be higher than a
 few MeV for a successful big bang nucleosynthesis~\cite{Kawasaki:2000en,Hannestad:2004px,Ichikawa:2005vw}.
Thus, the maximum temperature after inflation is estimated as
\begin{equation}
 T_{\rm max} \simeq \left( \sqrt{\frac{90}{\pi^2 g_*(T_R)}} T_R^2 M_P H_f \right)^{1/4} ,
\label{eq:Tmax}
\end{equation}
 where $H_f (\simeq H_+)$ is the Hubble parameter at the end of inflation.
In order for the thermal effect not to take the Higgs field back to the origin,
 one may impose 
\begin{equation}
 T_{\rm max} < v \,.
\label{cond:Tmaxbound}
\end{equation}
Otherwise, the relaxion can be destabilized and
 the tachyonic effective Higgs mass term can become negatively too large. 
One should notice, however, this condition implies
 that the sphaleron processes has been never activated
 after inflation.
This is also a serious trouble in the relaxion scenario
 from the viewpoint of baryogenesis.

\subsection{Experimental constraints on ALPs searches}

In literature, the constraints on ALPs have been displayed on the mass-coupling plane.
A relevant coupling is the photon-ALPs coupling normalized as
\begin{eqnarray}
 {\cal L} \supset -\frac{1}{4} g_{\gamma} \phi \tilde{F} F ,
\end{eqnarray}
 where $F$ and $\tilde{F}$ denote the field strength and its dual
 of the electromagnetic field.

In the relaxion model, another $\phi-\gamma-\gamma$ interaction is induced through
 the Higgs-relaxion mixing with the angle (\ref{mix:h-phi}).
This imposes a constraint on the mixing angle, which is
 proportional to $gM^2$.
In the rest of analysis, we will assume this condition is satisfied.

\subsection{Constraints summary}

The conditions to realize successful inflation and not to destabilize
 the relaxion field are expressed by Eqs.~(\ref{cond:smoothinflation}) and
 (\ref{cond:axion_stabil2}).
Experimental limits for ALPs search results are read
 for $g_{\gamma} =\alpha_{\rm em}/(2 \pi f)$.

Since constraints on the relaxion relic abundance (\ref{cond:omegah2-phi}) and
 reheating temperature (\ref{cond:Tmaxbound}) depend on not only the relaxion mass $m_{\phi}$ and decay constant $f$ but also $T_R$ and $N_{\rm osc}$,
 Eqs.~(\ref{cond:omegah2-phi}) and (\ref{cond:Tmaxbound})
 can be satisfied by those other parameters.
Hence, those constrains do not restrict parameter space on $m_{\phi}$ and $f$.

Using Eqs.~(\ref{cond:smoothinflation}) and (\ref{cond:axion_stabil2})
 we obtain
\begin{equation}
3m_\phi^2M_P^2 \gg v^4, \qquad  f^2 \ll 3M_P^2,
\end{equation} 
which implies $m_\phi \gg 10^{-5}$ eV and 
$g_\gamma \gg \alpha_{\rm em}/(2\sqrt3 \pi M_P) \sim 3 \times 10^{-22}$ GeV${}^{-1}$, 
independently of the Hubble scale. Meanwhile, the ALP mass has upper bound, $m_\phi \lesssim 10^4$ eV from 
astrophysical observations (see e.g., Ref.~\cite{Ringwald:2012hr,Arias:2012az,Graham:2015ouw}).
Therefore, the available region is obtained
 by $10^{-5}~{\rm eV} \ll m_\phi  \lesssim 10^4~{\rm eV}$.

The experiments exclude the region $ g_{\gamma} > 10 ^{-10}~{\rm GeV}^{-1}$
 for wide mass range.
In addition, from Eq.~(\ref{cond:smoothinflation}) we obtain the constraint,
\begin{equation}
 \log g_{\gamma}[{\rm GeV}]^{-1} = \log \alpha_{\rm em}/(2 \pi f)[{\rm GeV}]^{-1} \gg 
 \log m_\phi[{\rm eV}] -\log H[{\rm eV}] -20.
 \end{equation}
This constraint depends on the Hubble scale.

\begin{figure}[t]
 \begin{minipage}{0.49\hsize}
  \begin{center}
\includegraphics[width=80mm]{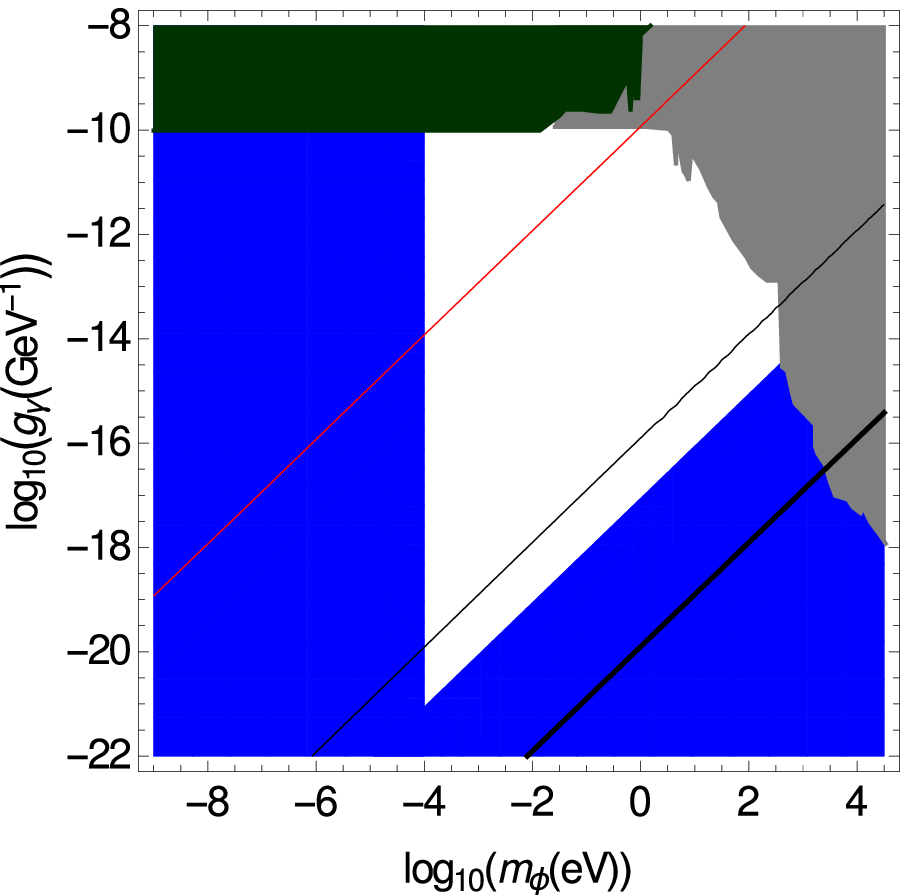}
  \end{center}
\end{minipage}
\hfill
 \begin{minipage}{0.49\hsize} 
  \begin{center}
\includegraphics[width=80mm]{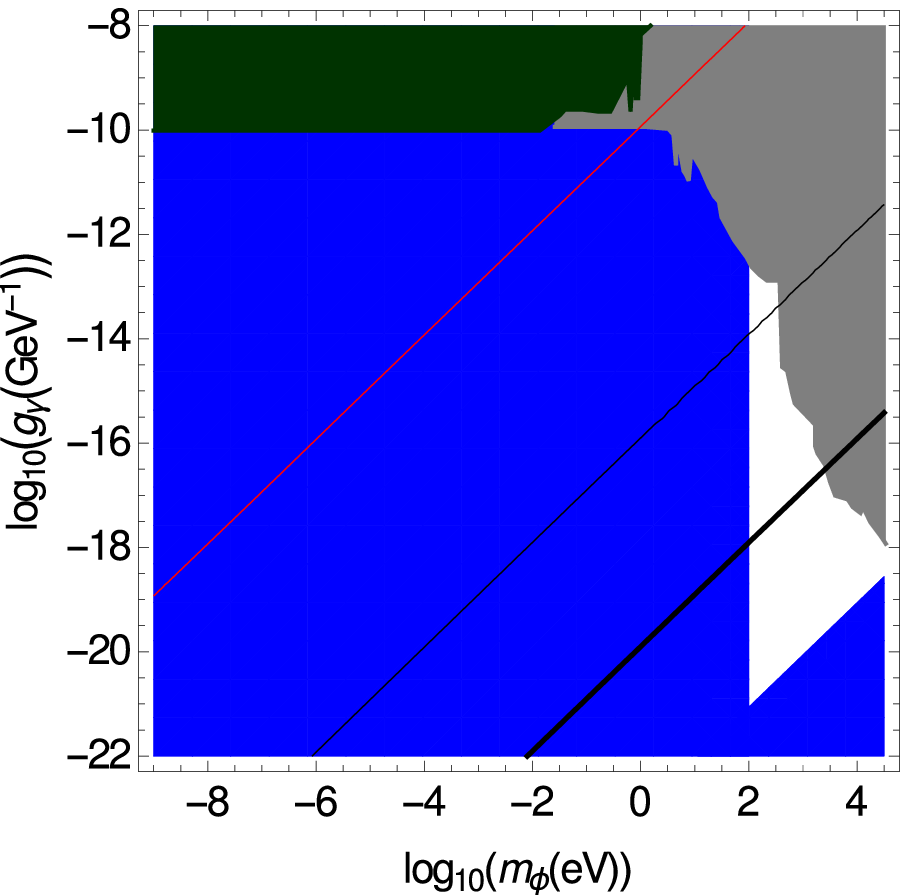}
  \end{center}
\end{minipage}
\caption{
The blue shaded region is excluded by the conditions~(\ref{cond:smoothinflation}) and
 (\ref{cond:axion_stabil2}). 
The dark green shaded region is the limit by ALPs search experiments. 
The grey shaded region is the limit by ALPs search experiments. We
 choose $H$ as $H=10^{-4}$ eV (left) and $H=10^2$ eV (right).}
\label{Fig:1}
\end{figure}

Figure~\ref{Fig:1} shows the viable region for $H=10^{-4}$ eV and $H=10^2$ eV.
The dark green shaded upper left region is excluded by the experiments,
mostly CAST helioscope experiment~\cite{Arik:2008mq}. The gray shaded
right region is excluded by the astrophysical observation and/or
astrophysical arguments. Those two regions are taken from Fig.~2
 in Ref.~\cite{Ringwald:2012hr}.
The left and lower blue shaded regions are excluded by
Eqs.~(\ref{cond:smoothinflation}) and (\ref{cond:axion_stabil2}). 
The red, black and thick oblique lines from the left to the right are
the contours for $\sqrt{m_{\phi} f}= 0.1$ GeV, $100$ GeV and $10$ TeV,
 which are about the scale of non-QCD strong dynamics $\Lambda $. 

As is clearly seen in Fig.~\ref{Fig:1}, the consistent value of $\Lambda$
 depends on the Hubble scale during inflation $H$.
From the left panel, for the lowest $H$, we need the strong dynamics scale below ${\cal O}(100)$ GeV
 with extra matter fields coupled with the Higgs fields to generate Higgs dependent coefficient of
 non-perturbative cosine potential like (\ref{pot:cos}) and  (\ref{eq:Lambda-Phi}).
That looks hardly possible.
As $H$ increases, the new dynamical scale increases.
As in its right panel, for example for $H=100$ eV,
 the strong scale can lie $100 $ GeV $\lesssim \Lambda \lesssim 100$
 TeV.

\begin{figure}[!t]
\begin{center}
\includegraphics[width=90mm]{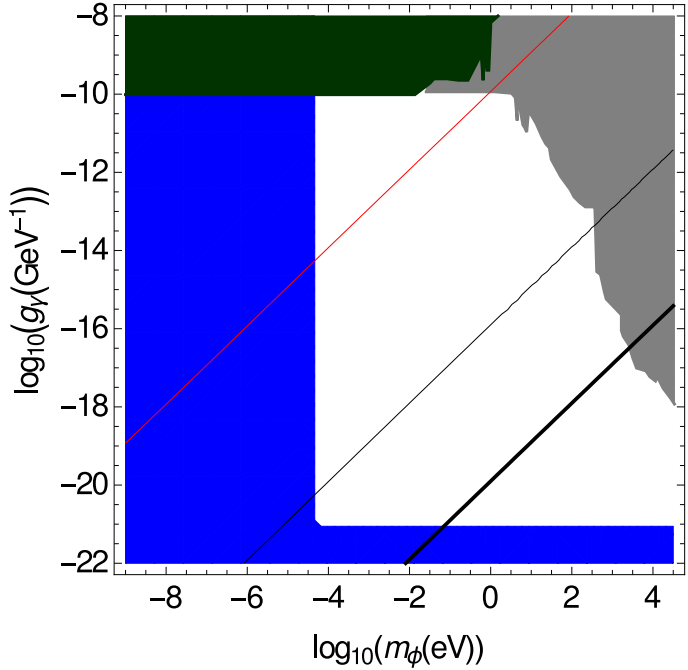}
\end{center}
\caption{The shaded regions  show the same constraints as those in Fig.~\ref{Fig:1}. 
Here, we leave the Hubble parameter as a free parameter, letting $H$
 scan all the cosmologically possible values. The uncolored region
 indicates the whole possible parameter space of the relaxion mass and the coupling to photon.}
\label{Fig:2}
\end{figure}

Figure~\ref{Fig:2} shows the whole consistent parameter space of the
relaxion mass and the coupling range for all the cosmologically possible
values of $H$. This provides a ``relaxion window'' based only
on cosmological argument, where the difficulty of a theoretical model
construction as mentioned above is not taken into account at all. 
It is interesting that there is the lower bound on $g_{\gamma}$, or 
equivalently the upper bound on the relaxion decay constant $f$.
%
%

\section{Concluding remarks}   \label{conclusion}

We have re-examined the cosmological constraints on the original relaxion
scenario proposed in Ref.~\cite{Graham:2015cka}.

First, we gave a caveat that depending on the sign of the terms in the
scalar potential, the QCD axion cannot play the role of the relaxion
even if there exists the coupling between the inflaton and the relaxion.

Second, we have shown that 
 the relic abundance of relaxion generated by the misalignment mechanism does not exceed the
 present dark matter density for sufficiently large values of $N_{\rm osc}$ or
 small potential height $m_{\phi}^2f^2$.

Third, 
 we derived ``the relaxion window'' of the cosmologically consistent mass and the
 coupling to the photon. Namely, we obtained 
 $10^{-5}$ eV $\ll m_{\phi} \lesssim 10^4$ eV and $3\times 10^{-22}~{\rm GeV}^{-1} \ll 
g_\gamma \lesssim 10^{-10}~{\rm GeV}^{-1}$.
The lower bound comes from the condition to stabilize the relaxion at a suitable vacuum
 with the Higgs field VEV and its mass of the weak scale. 
There is a strong correlation between $H$, in other words $m_{\phi}$, and the scale of
 strong dynamics $\Lambda$.
For higher mass, larger scale $\Lambda$ of non-QCD strong dynamics becomes available.
While for a low mass and larger $g_\gamma$,
 $\Lambda$ of non-QCD strong dynamics should be small and we need
 to construct a tricky model where non-QCD fermions are still somehow coupled
 with the SM Higgs field to generate the non-perturbative potential with
 the SM Higgs dependence. This is indeed challenging.

Another remaining issue is to build an inflation model which
 generates the observed primordial density perturbation. 
In addition, the electroweak symmetry is assumed not to be recovered after inflation
 in the relaxion scenario. 
Since the electroweak sphaleron mechanism has never been active, 
 it is necessary to contrive a viable baryogenesis mechanism.


\section*{Acknowledgments}
This project was completed under the support of Building of Consortia for
 the Development of Human Resources in Science and Technology.
This work is supported in part by the Grant-in-Aid for Scientific Research 
 No.~25400252 and 26247042 (T.K), No.~26400243 and No.~26105514 (O.S), No.~15K17654 (T.S)
 and No.~6887018 (Y.U) and from the Ministry of Education, Culture, Sports,
 Science and Technology in Japan. 
The work of O.S. is supported in part by the SUHARA Memorial Foundation.

%



\end{document}